\documentclass[a4paper,12pt]{spieman}  % use this instead for A4 paper
\usepackage{graphicx, siunitx}
\usepackage{setspace}
\usepackage{tocloft}
\usepackage{sidecap,booktabs}
\usepackage{enumitem}
\usepackage{xfrac}
\usepackage[T1]{fontenc}
\title{Evaluation of ultrasound sensors for transcranial photoacoustic sensing and imaging}
\author[ a,*]{Thomas Kirchner}
\author[ a,b]{Claus Villringer}
\author[ a]{Jan Laufer}
\affil[a ]{Fachgruppe Medizinische Physik, Institut für Physik, Martin-Luther-Universität Halle-Wittenberg, Germany}
\affil[b ]{Technische Hochschule Wildau, Germany}

\cftpagenumbersoff{figure}
\cftpagenumbersoff{table} 
\begin{document} 
\maketitle

\begin{abstract}

\noindent \textbf{Significance:} Biomedical photoacoustic (PA) imaging is typically used to exploit absorption-based contrast in soft tissue at depths of several centimeters. When it is applied to measuring PA waves generated in the brain, the acoustic properties of the skull bone cause not only strong attenuation but also a distortion of the wavefront, which diminishes image resolution and contrast. This effect is directly proportional to bone thickness. As a result, transcranial PA imaging in humans has been challenging to demonstrate.

\noindent \textbf{Aim:} We measured the acoustic constraints imposed by the human skull to design an ultrasound sensor suitable for transcranial PA imaging and sensing.

\noindent \textbf{Approach:} We calculated the frequency dependent losses of human cranial bones \emph{in silico} and performed measurements \emph{ex vivo} using broadband ultrasound sources based on PA excitation, such as a single vessel phantom with tissue-mimicking optical absorption. We imaged the phantoms using a planar Fabry-Perot sensor and employed a range of piezoelectric and optical ultrasound sensors to measure the frequency dependent acoustic transmission through human cranial bone.

\noindent \textbf{Results:} Transcranial PA images show typical frequency and thickness dependent attenuation and aberration effects associated with acoustic propagation through bone. The skull insertion loss measurements showed significant transmission at low frequencies. In comparison to conventional piezoelectric sensors, the performance of plano-concave optical resonator (PCOR) ultrasound sensors was found to be highly suitable for transcranial PA measurements. They possess high acoustic sensitivity at a low acoustic frequency range that coincides with the transmission window of human skull bone. PCOR sensors showed low noise equivalent pressures and flat frequency response which enabled them to outperform conventional piezoelectric transducers in transcranial PA sensing experiments.

\noindent \textbf{Conclusions:} Transcranial PA sensing and imaging requires ultrasound sensors with high sensitivity at low acoustic frequencies, and a broad and ideally uniform frequency response. We designed and fabricated PCOR sensors and demonstrated their suitability for transcranial PA sensing.

\end{abstract}

% Include a list of up to six keywords after the abstract

% Include email contact information for corresponding author
{\noindent \footnotesize\textbf{*} corresponding author:\linkable{thomas.kirchner@physik.uni-halle.de}}

\begin{spacing}{1}

\section{Introduction}
\label{sec:intro}
Biomedical photoacoustic (PA) monitoring and imaging of the brain through the intact skin and skull is hampered by the effects of the acoustic propagation through the bone tissue. Cranial bones cause strong attenuation and wavefront distortion, and thus diminish image resolution and contrast. Previous studies of PA wave propagation through cranial bone often relied on numerical simulations where spatial distribution of the speed of sound and mass density of the bone tissue were estimated from x-ray computer tomography (CT) data. This typically involved the calculation a porosity map \cite{aubryExperimentalDemonstrationNoninvasive2003} from clinical resolution x-ray CT in Hounsfield units or by defining a simplified layered skull model \cite{liangAcousticImpactHuman2021}. However, the limited spatial resolution of clinical CT images does not allow the representation of fine internal microstructure in acoustic propagation models \cite{pintonAttenuationScatteringAbsorption2012} and introduces partial volume effects. Despite these limitations, conventional x-ray CT image data has been used to correct for wavefront aberrations in PA image reconstruction algorithms \cite{naTranscranialPhotoacousticComputed2020} or in clinical transcranial high intensity focused ultrasound (HIFU) applications \cite{marquetNoninvasiveTranscranialUltrasound2009}. More accurate simulations of transcranial acoustic wave propagation were demonstrated recently using high resolution x-ray micro-CT images \cite{robertsonEffectsImageHomogenisation2018}. The feasibility of monitoring the brain using the PA effect was demonstrated experimentally in small \cite{wangNoninvasiveLaserinducedPhotoacoustic2003, wang_noninvasive_2006} and large animals \cite{kirchnerPhotoacousticMonitoringBlood2019, kirchnerPhotoacousticsCanImage2019, petrovaNoninvasiveMonitoringCerebral2009}, and in humans who had undergone hemicraniectomy \cite{naMassivelyParallelFunctional2021}. PA measurements on the human brain \emph{in vivo} through the intact skin and skull would lead to clinical applications such as monitoring of brain hemodynamics, the detection of intracranial bleeding and stroke diagnosis. However, while transcranial PA imaging through human skull bone has been investigated in phantoms or using \emph{ex vivo} primate skulls in water baths \cite{huangAberrationCorrectionTranscranial2012}, PA measurements in humans through the skull bone have yet to be demonstrated \emph{in vivo}.
The main challenge for human transcranial PA measurements lies in the acoustic attenuation and wavefront aberration in human cranial bone. \cite{yaoPhotoacousticBrainImaging2014, liangAcousticImpactHuman2021} This effect is highly frequency dependent \cite{leeAttenuationHumanSkull2020} and dependent on bone thickness. Since the initial pressure in the brain is limited by the maximum permissible exposure on the skin and the optical attenuation by the skin, bone, and brain tissues, it is vital to develop ultrasound (US) sensors that are optimised for this application. Previous studies have typically relied on piezoelectric sensors with large active element sizes as they are readily available and are widely regarded as the most sensitive \cite{winklerNoiseequivalentSensitivityPhotoacoustics2013, xiaOptimizedUltrasoundDetector2013}. While highly sensitive optical US sensors have recently been reported~\cite{guggenheimUltrasensitivePlanoconcaveOptical2017}, these technologies have not yet been employed for transcranial PA measurements.

In this paper, we investigate the feasibility of PA sensing and imaging of the human brain through the intact skull. We simulate the frequency dependent attenuation of broadband PA waves in human cranial bone and compare the \emph{in silico} results with measurements made in \emph{ex vivo} human skull bone using piezoelectric and optical US sensors. We fabricated and characterized plano-concave optical resonator (PCOR) sensors based on the design first proposed by Guggenheim et al.~\cite{guggenheimUltrasensitivePlanoconcaveOptical2017}. By changing the material composition of the PCOR sensors, the acoustic sensitivity was increased. PA transmission measurements through \emph{ex vivo} human skull bone showed that PCOR sensors offer greater acoustic sensitivity compared to large-area piezoelectric transducers, making them highly suited to transcranial PA measurements. 

\section{Materials and methods}
\subsection{Human skull sample}
The skull of a 70 year old male body donor was used for all measurements. Written informed consent for general scientific investigation was given by the body donor prior to death. A high resolution x-ray micro-CT scan and bone segmentation of the skull (see data set and descriptor\cite{kirchnerMicroCTHumanSkull2022}) was the basis for numerical simulations of transcranial acoustic propagation.
PA transmission measurements and PA imaging were performed across three location on the skull, i.e., across temporal, occipital and frontal cranial bone as indicated in figure \ref{fig:fabry}b. The skull bone at these sites differs in thickness and porosity. The temporal bone is thinnest and composed of the least porous tissue while the cranial and frontal bone represent the thickest and most porous regions. The skull sample was immersed in degassed and deionised water at room temperature between 21 and \SI{22}{\celsius} for at least 15 minutes before the measurements. 

\subsection{Simulation of transcranial acoustic propagation}
3D PA wave propagation through frontal cranial bone was simulated using k-Wave\cite{treeby_k-wave:_2010}. The PA source was represented by a disk of \SI{5}{\milli\meter} diameter and a single-voxel thickness of \SI{125}{\micro\meter}. The acoustic sensor was placed at a distance of \SI{3}{cm} from the source, parallel to the source and on the same acoustic axis. We defined a two medium volume based on the \SI{125}{\micro\meter} resolution bone segmentation in the micro-CT data set \cite{kirchnerMicroCTHumanSkull2022} to represent the frontal cranial bone. The acoustic properties were set to those of bone with an acoustic attenuation $\alpha$\,=\,\SI{2.7}{db\per\cm\per\mega\hertz\squared}, a mass density $\rho_\textrm{bone}$\,=\,\SI{2190}{\kilo\gram\per\meter\cubed} and a speed of sound $c_\textrm{bone}$\,=\,\SI{3100}{\meter\per\second}, and those of water with a mass density $\rho_\textrm{water}$\,=\,\SI{1000}{\kilo\gram\per\meter\cubed}, speed of sound $c_\textrm{water}$\,=\,\SI{1480}{\meter\per\second} and negligible acoustic attenuation. Acoustic wave propagation was modelled in water as a reference and frontal cranial bone in a water bath. The power density spectra of the signals were calculated using the Fourier transform, from which the bone insertion loss was obtained. 

Only longitudinal pressure waves were simulated since shear waves are much less prominent and even negligible at low angles of incidence \cite{whiteLongitudinalShearMode2006}. Normal incidence was chosen to minimise the overall attenuation.

\subsection{Transcranial PA imaging using a Fabry-Perot tomograph} \label{Transcranial PA imaging using a Fabry-Perot tomograph}
\begin{figure}[hbt]
\centering
\includegraphics{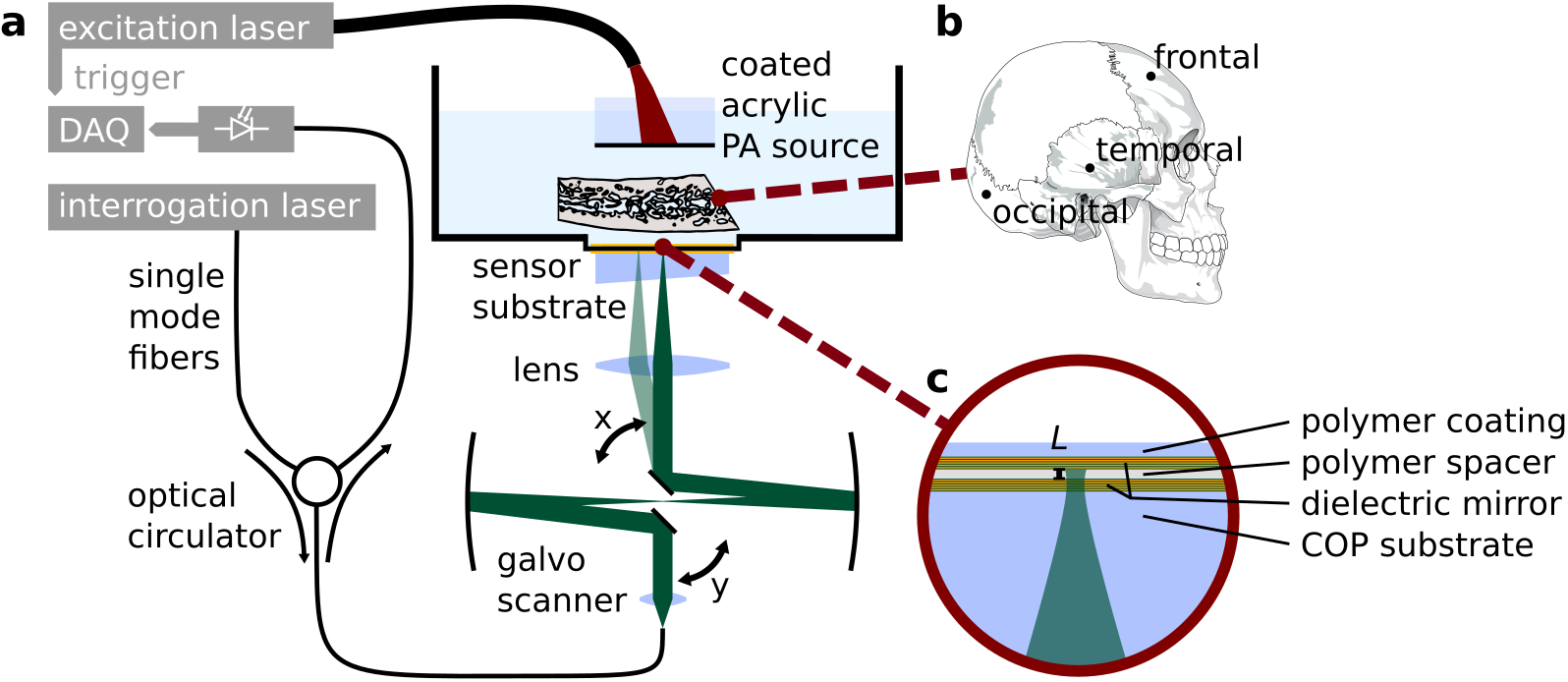}
\caption[Fabry-Perot interrogation setup]{\label{fig:fabry} Fabry-Perot raster scanning setup for transcranial tomographic imaging of a planar PA source and a tube phantom. (\textbf{a}) Experimental setup for transmission measurements through \emph{ex vivo} human skull. Target absorbers, such as a thin film of acrylic black paint on a PMMA substrate (shown here) or a tube filled with a liquid absorber, were illuminated by the output of an excitation laser and generated PA fields, these propagated through a region of cranial bone (see b) and were measured by scanning a interrogation laser spot over a planar Fabry-Perot sensor.  (\textbf{b}) Diagram of a human skull and the measurement locations. (\textbf{c}) Schematic of the planar Fabry-Perot sensor.}
\end{figure}

To investigate the effects of transcranial acoustic wave propagation on PA images, simple absorbing structures were imaged though human cranial bone using a Fabry-Perot raster scanning tomograph (shown in figure \ref{fig:fabry}a). The scanner is based on Zhang et al.~\cite{zhangBackwardmodeMultiwavelengthPhotoacoustic2008} and incorporates a planar Fabry-Perot (PFP) sensor described in Buchmann et al.~\cite{buchmannCharacterizationModelingFabryPerot2017}. The sensor consists of two dielectric mirrors separated by a parylene C spacer deposited on a cyclo-olefin polymer (COP) backing substrate as shown in figure \ref{fig:fabry}c. The physical thickness of the Fabry-Perot sensor ($L$\,=\,\SI{22}{\micro\meter}) results in a near constant acoustic frequency response ranging from dc to 38\,MHz (-3\,dB). The transduction mechanism of the sensor is based on the detection of acoustically induced changes in the reflected optical power of a cw interrogation beam following the pulsed excitation of PA signals in the target. 3D PA image data sets were acquired by 2D raster scanning the interrogation beam across the detection aperture. 

The cw output of an interrogation laser (T100S, Yenista Optics, Lannion, France) was focused on the Fabry-Perot sensor with a beam waist radius of \SI{30}{\micro\meter}. The beam is raster-scanned across an aperture of 4\,cm$^2$ using galvanometer mirrors (GVS012, Thorlabs) and the reflected intensity is coupled to an InGaAs photodiode (G9801-22, Hamamatsu Photonics K.K., Japan) using an optical circulator. The acquisition of PA signals at each point was triggered by an Nd:YAG excitation laser (Nano L 150-50, Litron Lasers, Rugby, UK) which provided excitation pulses at 1064\,nm with a duration of 7 to 9\,ns and at a pulse repetition rate of 50\,Hz. The photodiode output was high-pass filtered at 50\,kHz and recorded using a digital oscilloscope (PCI-5124, National Instruments, USA) at a sampling rate of 100\,MS/s. The water tank above the sensor was filled with degassed and deionised water in which the skull samples and PA sources were immersed.

PA images of two absorbing structures were acquired. The first structure consisted of a thin layer of black acrylic paint on a planar PMMA substrate positioned on the same acoustic axis and in parallel to the sensor at a distance of 3\,cm. The layer of paint was illuminated directly by the divergent output of a multimode fiber. The pulse energy was 2\,mJ and the spot size diameter was 5\,mm, resulting in a planar PA wave with broadband acoustic frequency content. The excitation pulses were attenuated to 5\% for the reference measurement without skull. The second structure was intended to represent a vascular target and consisted of a silicone tube (2\,mm inner diameter) filled with a \SI{2.2}{M} aqueous nickel sulfate solution (Sigma-Aldrich, CAS 10101-97-0) in a water bath. The tube was positioned at a distance of 2\,cm from the sensor and was illuminated directly with 30\,mJ pulses, resulting in a fluence of $\approx$\,\SI{100}{mJ\per\centi\meter\squared}. The absorption coefficient of the solution at 1064\,nm was $\mu_\mathrm{a}$\,=\,\SI{6}{\per\centi\meter} -- comparable to the optical absorption of whole blood in the near infrared window. The solution is likely to exhibit a higher Grüneisen coefficient compared to that of blood\cite{fonsecaSulfatesChromophoresMultiwavelength2017}. PA image data sets were acquired in water as a reference measurement, and through skull bone by positioning the temporal, occipital, and frontal cranial bone between the PA source and the sensor. The PA image volumes are reconstructed with a fast Fourier transform algorithm\cite{kostliTemporalBackwardProjection2001} for a planar sensor using the k-Wave toolkit\cite{treeby_k-wave:_2010}.

\subsection{PA measurements of skull insertion loss with piezoelectric and optical US sensors}
To investigate the effect of the type of US sensor on the sensitivity of transcranial PA measurements, the acoustic transmission through frontal cranial bone was measured using single element piezoelectric and optical sensors. PA waves were again generated in the black acrylic coating on a PMMA substrate, and the sensors were positioned in parallel to the PA source, at a distance of 3\,cm and on the same acoustic axis.

A total of three piezoelectric sensors were investigated. They included two unfocused lead zirconium titanate (PZT) transducers with an active element diameter of 25.4\,mm, a bandwidth of 65\,\% relative -6\,db at a center frequencies of either 500\,kHz (V301-SU, Olympus, Waltham, USA) or 1\,MHz (V302-SU, Olympus, Waltham, USA), and a broadband polyvinylidene fluoride (PVDF) sensor (Precision Acoustics Ltd, Dorchester, UK) with -6\,db at 20\,MHz and an active element diameter of 19.2\,mm. PA signals were were recorded using a data acquisition unit with a sampling rate of 80\,MHz (Flash ADC, PhotoSound, Houston, USA). The optical US sensors included the planar Fabry-Perot sensor used in the tomograph and a set of plano-concave optical resonator (PCOR) sensors, the fabrication and characterisation of which is described in the following section. Piezoelectric sensors were used to measure 2500 waveforms. Optical sensor measurements were performed with a $L$\,=\,\SI{22}{\micro\meter} planar Fabry-Perot sensor (recording 100 waveforms, each pre-averaged over 25 acquisitions) and with the largest PCOR sensors ($L$\,=\,\SI{493}{\micro\meter}). The optical sensors were interrogated with a beam of \SI{30}{\micro\meter} waist radius.

\subsection{Fabrication and characterisation of plano-concave optical resonator (PCOR) sensors}
\begin{figure}[hbt]
\centering
\includegraphics{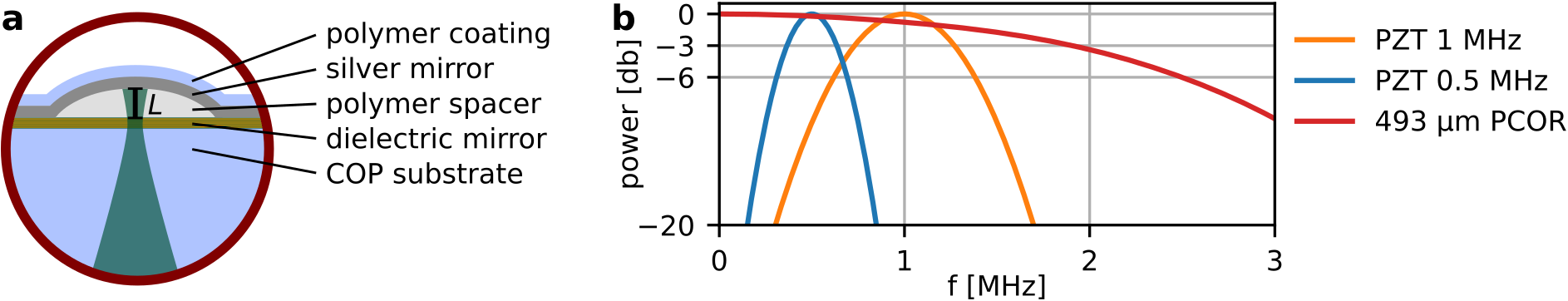}
\caption[PCOR layout and sensitivity]{\label{fig:pcortheory} (\textbf{a}) Schematic of a plano-concave optical resonator (PCOR) sensor with a physical spacer thickness $L$. (\textbf{b}) Simulated acoustic sensitivity spectra of a PCOR sensor \cite{beardTransductionMechanismsFabryPerot1999} with $L$\,=\,\SI{493}{\micro\meter} and two PZT-based, piezoelectric US transducers (frequency responses with a fractional bandwidth of 65\,\%).}
\end{figure}

Figure \ref{fig:pcortheory}a shows a cross-section of the PCOR sensors. Planar substrates of cyclo-olefin polymer (COP) material (ZEONEX 480R, Zeon, Chiyoda, Japan) were made using injection molding (Polymeroptix, Goch, Germany). The substrates have a thickness of 9\,mm. Dielectric mirrors consisting of SiO$_2$ and TiO$_2$ were deposited using sputtering. The reflectivity of the mirrors was either 94.9\,\% or 97.8\,\% at 1580\,nm. Using an inkjet dispenser (AL 300, ficonTEC, Achim, Germany), droplets of a UV-curable, liquid polymer (OrmoClad, Micro Resist Technology, Berlin, Germany) were deposited on the first mirror. The droplets were varied in size (ranging from 100 to \SI{500}{\micro\meter}) to provide sensors with different acoustic sensitivity and frequency response, and were cured using ultraviolet (UV) light in a two-stage process. The  droplets were illuminated immediately after deposition with 365\,nm light (UV-LED Solo P, Opsytec, Ettlingen, Germany) at a fluence of $\approx$\,\SI{4}{W\per\centi\meter\squared} for 10\,s. The substrates were then placed in a chamber flushed pure nitrogen to counter the  inhibitory effect of oxygen on UV curing \cite{alvankarianExploitingOxygenInhibitory2015} and again illuminated by UV light for at least 30\,s. The polymer spacer material was chosen because it has an exceptionally low optical absorption in the 1530--1625\,nm wavelength band. The second mirror consists of a 60\,nm silver layer with a reflectivity of 97\,\% and was deposited using DC magnetron sputtering (EM SCD 500 Leica, Wetzlar, Germany). A final protective barrier coating (\SI{10}{\micro\meter}, parylene C) was deposited to prevent water damage.

PCOR sensors have been shown to provide ultrahigh sensitivity and broad frequency res\-ponse~\cite{guggenheimUltrasensitivePlanoconcaveOptical2017}. The latter is illustrated in Figure \ref{fig:pcortheory}b, which shows the theoretical frequency responses of a comparatively thick PCOR sensor (spacer thickness of $L$ = \SI{493}{\micro\meter}) in comparison with two of the PZT transducers used in this study. While the PZT transducers suffer from low sensitivity at low acoustic frequencies, PCOR sensors have a near uniform frequency response from dc to several MHz -- depending on their size.

%The substrates were either wedged to reduce interference within the COP substrate or a wedged glass window was fused to the passive side with optical adhesive. 

The noise equivalent pressure (NEP) is a commonly used metric for the acoustic sensitivity of a sensor \cite{winklerNoiseequivalentSensitivityPhotoacoustics2013} and is defined as the pressure at which the peak positive signal amplitude, $\textrm{max}(S)$, is equal to the standard deviation of the detection noise, $\sigma(\textrm{noise})$, i.e., $\textrm{NEP} = \sigma(\textrm{noise}) \cdot p_\textrm{pp} / \textrm{max}(S)$, where $p_\textrm{pp}$ is the peak positive pressure amplitude effected by a calibrated source. It can also be normalized over a specific detection bandwidth as NEP per $\sqrt{\textrm{bandwidth}}$.

The PCOR sensors were incorporated into the Fabry-Perot scanner described in section \ref{Transcranial PA imaging using a Fabry-Perot tomograph}. The sensors were operated using the same interrogation and data acquisition setup. The measurement of the acoustic sensitivity involved the use of a calibrated piezoelectric transducer (Sonotec, Halle, Germany) with an active element diameter of 20\,mm and a center frequency of 1\,MHz  as an acoustic source. The transducer was placed in a water tank at a distance of 20\,mm to the PCOR sensor and on the same acoustic axis. The sensor position within the sound field was chosen as it coincides with a local pressure maximum, thus minimising measurement bias toward underestimating NEP. The calibrated transducer was driven by a frequency generator (Agilent 33522A,  Santa Clara, USA) to produce a tone burst consisting of a four period sine wave with 100\,mVpp at a frequency of 1\,MHz resulting in a peak positive pressure amplitude of $p_\textrm{pp}$\,=\,230\,Pa at the location of the PCOR sensor. $\sigma(\textrm{noise})$ was measured over 500 data points preceding the tone burst. The NEP of each PCOR sensor was estimated from the measured acoustic signal. To allow a comparison of sensors of varying size and therefore acoustic detection bandwidths, a digital band-pass filter (2nd order Butterworth) was applied to the data. For each PCOR sensor, the pass-band ranged from 10\,kHz to the first minimum in sensitivity (cf.\ supplementary figure 1 and listed as \emph{cutoff frequency} in supplementary table 1). The NEP values correspond to the -3\,db bandwidth of each sensor, which was estimated using a numerical model \cite{beardTransductionMechanismsFabryPerot1999}. The PCOR sensor with the lowest NEP was used in the frontal cranial bone insertion loss experiments.

\section{Results and discussion}
\subsection{Transcranial ultrasound propagation in silico}

\begin{figure}[bht]
\centering
\includegraphics{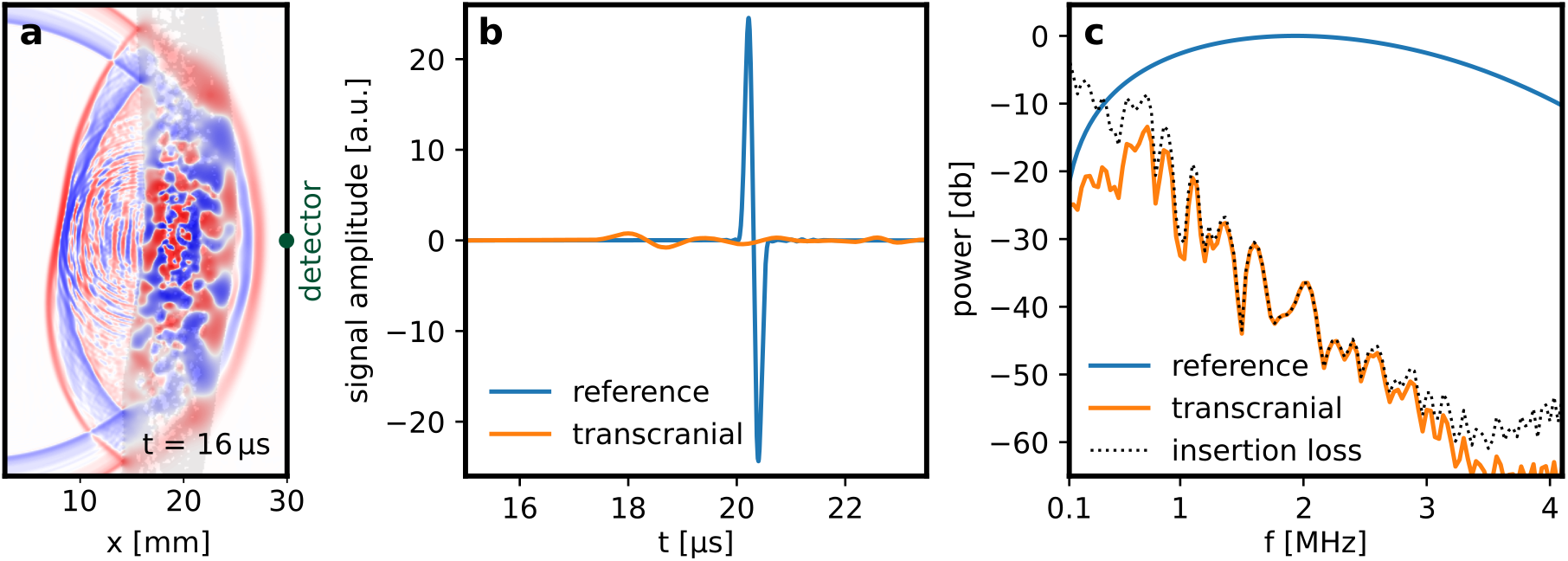}
\caption[transcranial PA in silico]{\label{fig:insilico} Acoustic propagation through frontal cranial bone \emph{in silico}. (\textbf{a}) Snapshot at time step $t$\,=\,\SI{16}{\micro\second} of the centre slice of a 3D acoustic field resulting from the propagation of a plane wave with \SI{5}{mm} diameter across frontal cranial bone (see also supplemental video 1). The light gray area shows the segmented skull tissue which was assigned the acoustic properties of bone during the simulation. The green dot indicates the detector position for signals and frequency spectra shown in (b,c) (\textbf{b}) Single waveforms detected at center voxel with distance $x$\,=\,\SI{30}{mm} from the acoustic source. Simulation results without skull shown as reference. (\textbf{c}) Acoustic power spectrum of the detected waveforms and the calculated insertion loss. 
}
\end{figure}

The numerical simulation of the ultrasound propagation through human frontal cranial bone is illustrated in figure \ref{fig:insilico}a, which shows a 2D snapshot of the acoustic field \SI{16}{\micro\second} after the generation of a PA plane wave. A video of the wave propagation is provided in supplementary video 1. 

Figure \ref{fig:insilico}b shows the waveforms acquired by a point detector on the acoustic axis at a distance of $x$\,=\,\SI{3}{cm} from the planar acoustic source. The reference signal (blue line) corresponds to simulated acoustic propagation in water while the transcranial signal (orange line) shows the effects of transmission through cranial bone, i.e., strong attenuation of the transcranial waveform with respect to the reference, as well as additional reverberations later in the signal. The attenuation is mainly caused by the multiple reflections within the cancellous bone due to acoustic impedance mismatch and less by the acoustic attenuation within the solid bone. This was verified by conducting additional simulations with attenuation coefficients $\alpha$ ranging from \SI{2.7} to \SI{6.0}{db\per\centi\meter\per\mega\hertz\squared}, and was found to have  a negligible effect on the overall attenuation.

Figure \ref{fig:insilico}c shows the simulated frequency dependent acoustic insertion loss calculated from the the acoustic power spectra of the transcranial and reference waveforms. The spectrum of the insertion loss shows a reduction in the transmitted acoustic power of approximately -20\,db at 1\,MHz. The insertion loss increases with frequency at a rate of around -10\,dB/MHz. Figure \ref{fig:insilico}c indicates that PA measurements through the frontal cranial bone may be feasible at frequencies below 2 \,MHz.

\subsection{Ex vivo transcranial PA tomography using a planar Fabry-Perot sensor}
\begin{figure}[hbt]
\centering
\includegraphics{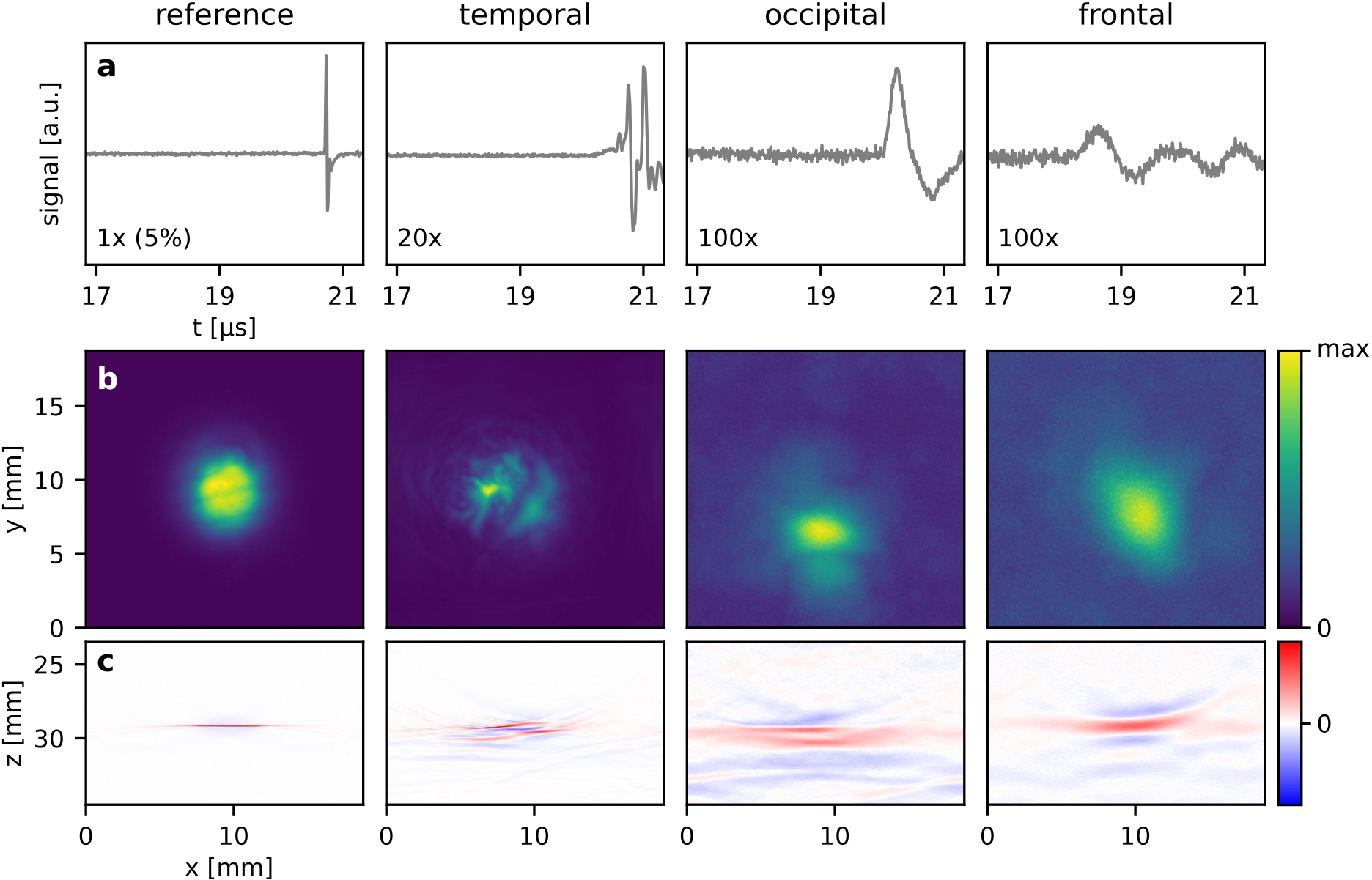}
\caption[transcranial PAI ex vivo planar]{\label{fig:planar} PA imaging of a layer of black paint on a PMMA substrate source using planar Fabry-Perot sensor. The left column shows the reference measurement, the columns to the right show PA measurements through temporal, occipital and frontal cranial bone. (\textbf{a}) Representative PA waveforms acquired at a single detection point. (\textbf{b}) \textit{x--y} maximum intensity projections (MIP) of the reconstructed PA images shown in a linear color scale. (\textbf{c}) Central \textit{x--z} slice shown in a diverging linear color scale.}
\end{figure}

Figure \ref{fig:planar} shows the image data sets measured in a planar absorber. The PA source had a diameter of 5\,mm and the area of the detection aperture was \SI{4}{cm\squared}. In addition to the reference measurement in water, we inserted temporal, occipital or frontal cranial bone between the source and the sensor. Representative PA waveforms measured at a single point through water (reference) and the different types of bone tissue are shown in figure \ref{fig:planar}a. As expected, the attenuation is lowest for the comparatively thin and least cancellous temporal bone while an attenuation of more than two orders of magnitude was observed in thicker and more cancellous occipital and frontal cranial bone. 

\begin{figure}[hbt]
\centering
\includegraphics{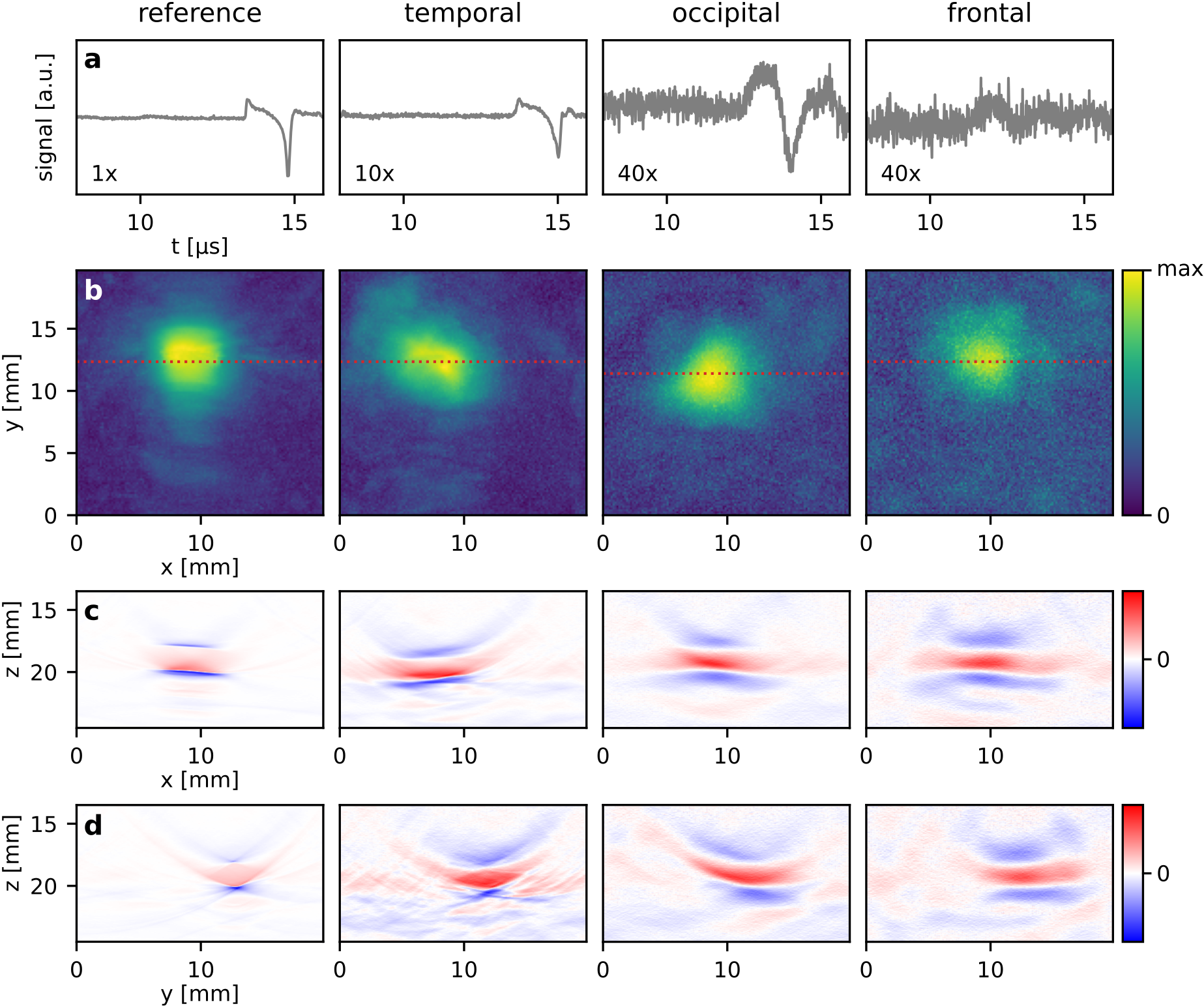}
\caption[transcranial PAI ex vivo tube]{\label{fig:tube} PA imaging of a 2\,mm inner diameter tube filled with a 2.2\,M NiSO$_4$ solution using a planar Fabry-Perot sensor. The left column shows the reference measurement, the columns to the right show the measurements through temporal, occipital, and frontal cranial bone. (\textbf{a}) Representative PA waveforms acquired at a single detection point. (\textbf{b}) $x$--$y$ MIPs of the reconstructed PA images shown in linear color scale. (\textbf{c}) Selected $x$--$z$ longitudinal slices along the red dotted lines in (b) shown in a linear color scale. (\textbf{d}) Selected $y$--$z$ axial slices at $x$\,=\,9\,mm.
}
\end{figure}

Maximum intensity projections (MIP) of the reconstructed 3D images are shown in figure\,\ref{fig:planar}b and central slices in in figure\,\ref{fig:planar}c. A comparison of the MIPs of the reference measurement and that for temporal bone show that the transmission of PA waves through a comparatively thin and least porous bone results in significant aberrations as evidenced by the distorted visualisation of the planar PA source in the reconstructed images. The effects of wavefront aberrations and frequency dependent attenuation are even more significant in occipital and frontal cranial bone. While the location of the planar PA source can be discerned, its shape is distorted and smeared across a much larger volume. The resolution we observed in transcranial MIPs is on the order of 1\,mm. It should be noted that we assumed a homogeneous speed of sound in the reconstruction algorithm. Its value was varied depending on bone thickness. For the images in figures \ref{fig:planar}\ and \ref{fig:tube}, the assumed speed of sound ranged from \SI{1480}{\meter\per\second} for the reference measurement to \SI{1660}{\meter\per\second} for frontal cranial bone. While more advanced algorithms have been reported that use additional information for a more accurate reconstruction, such as the distribution of acoustic properties inferred from X-ray CTs\cite{naTranscranialPhotoacousticComputed2020}, the focus of this work is on the investigation of the capabilities of US sensors. Also, coregistered X-ray Micro-CT data would not be available in a clinical setting. We therefore opted for a simple heuristic reconstruction as it illustrates the challenges in transcranial PA imaging.

The effects of wavefront aberrations and attenuation in bone tissue are also seen in the image data sets acquired in the vascular target structure (figure \ref{fig:tube}). While the amplitude of the PA signal excited in the absorber-filled tube is attenuated by an order of magnitude after propagating through temporal bone (figure \ref{fig:tube}a), its shape is maintained. This suggests that its frequency content has not been strongly affected, which is confirmed by the MIPs and image slices of the reconstructed image volumes in figures \ref{fig:tube}c-d in which the image acquired through the temporal bone is, apart from minor distortions, comparable to the reference image. By contrast, the PA signals and images acquired through the thicker and porous occipital and frontal cranial bone tissue show strong attenuation and smearing of the image due to wavefront distortion. While the absorption coefficient of the absorbing solution (NiSO$_4$) is comparable to that of whole blood, the fluence used to excite PA waves is at least two orders of magnitudes higher than what one could optimistically expect to transmit into the brain in an \emph{in vivo} imaging scenario. Despite the high fluence in the tube, the PA signal measured through frontal cortical bone was dominated by noise. It is therefore doubtful that PA transcranial brain imaging could be achieved in humans \emph{in vivo} using the planar Fabry-Perot sensor employed in this study. This is not surprising since its capabilities are not well matched to the requirements of transcranial PA detection. While the planar Fabry-Perot sensor offers a broadband frequency response up to tens of MHz, transcranial measurements benefit from high acoustic sensitivity over a comparatively low frequency band (up to 2 \,MHz). Importantly, interferometric optical sensors can be designed to fulfill these requirements. Increasing the optical thickness of the resonator will not only reduce the frequency response of the sensor but also maximise its acoustic sensitivity -- an approach we pursued by developing plano-concave optical resonator sensors.

\begin{figure}[hbt]
\centering
\includegraphics{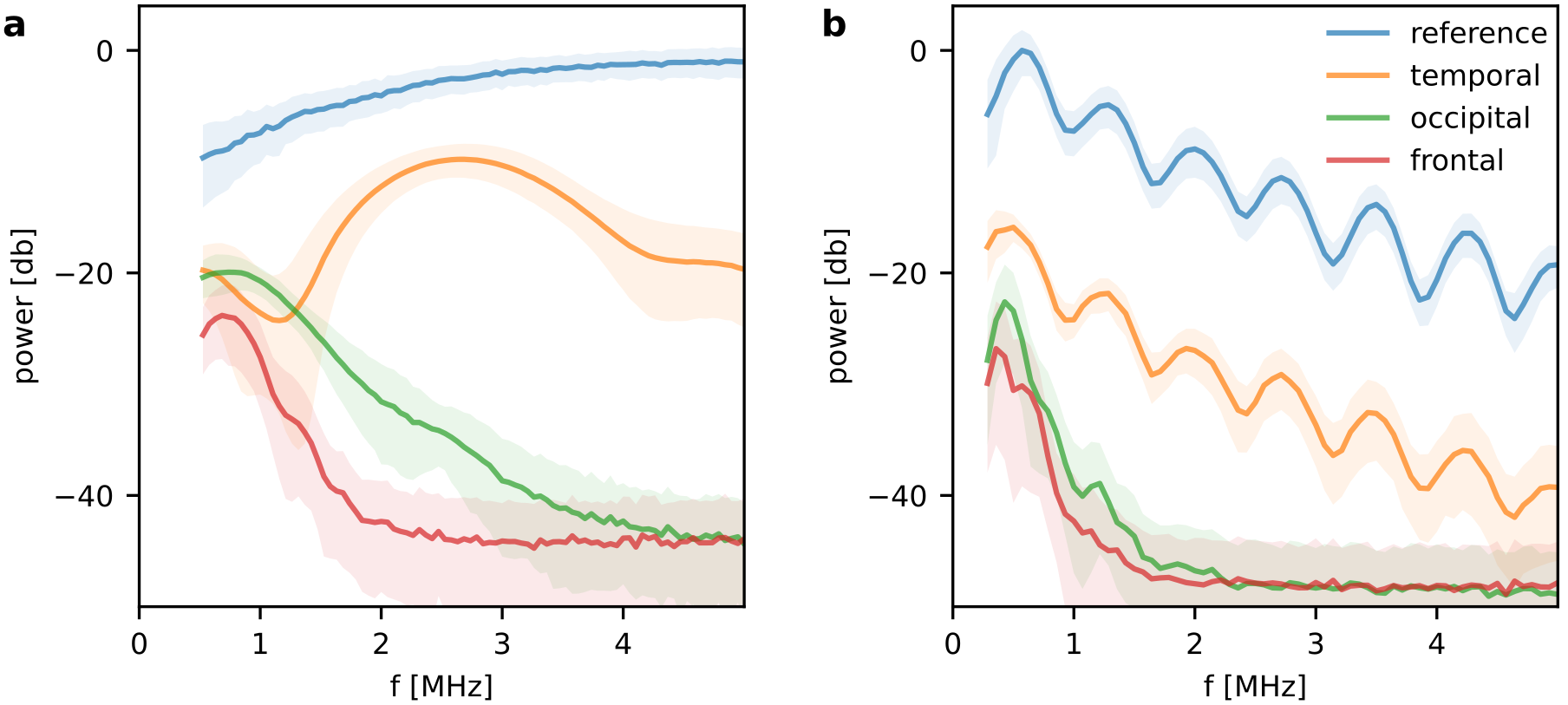}
\caption[transcranial PAI acoustic spectra]{\label{fig:paispectra} Acoustic power spectra of the image data sets measured in transcranial PA imaging experiments. The mean is shown as a solid line and the standard deviation is shown as a shaded region. (\textbf{a}) Power spectra of a planar broadband PA source (thin black coating on a PMMA block), and (\textbf{b}) a vascular target (silicone tube with a 2\,mm inner diameter, filled with a NiSO$_4$ solution).}
\end{figure}

The thicker and more porous the bone tissue, the stronger the effects of acoustic attenuation, aberration, and the loss of image resolution and blurring of features. The reduction in image resolution is caused by the frequency dependent attenuation of the bone tissue which is shown in figure \ref{fig:paispectra}a for the planar, broadband PA source. In the measurements through the temporal bone, higher acoustic frequencies are transmitted compared to the measurements made in occipital and frontal bone, which primarily transmit frequencies below 2\,MHz. The temporal bone is also called the transcranial window and used clinically for Doppler-US measurements on a sub-cranial artery using transducers with a center frequency around 2\,MHz. This trend is also confirmed by the measurements made in the tube phantom (\ref{fig:paispectra}b) albeit with a reference spectrum that is affected by the optical and material properties of the target. 

While the imaging data confirm that transcranial PA measurements and imaging is at least feasible, an important question is what effect of the choice of US sensor has on the sensitivity of the measurement. Given that bone attenuates acoustic frequencies above 1.5\,MHz strongly, an intuitive choice would be to select piezoelectric transducers with large active areas to maximise acoustic sensitivity while it is less clear if optical sensors could play a role. In the next section, we compare the sensitivity of low-frequency piezoelectric and optical US sensors for transcranial PA sensing through frontal cranial bone.

\subsection{Ex vivo transcranial PA measurements with low-frequency ultrasound sensors}  
\label{Ex vivo transcranial PA measurements with low-frequency ultrasound sensors}
\begin{figure}[tbh]
\centering
\includegraphics{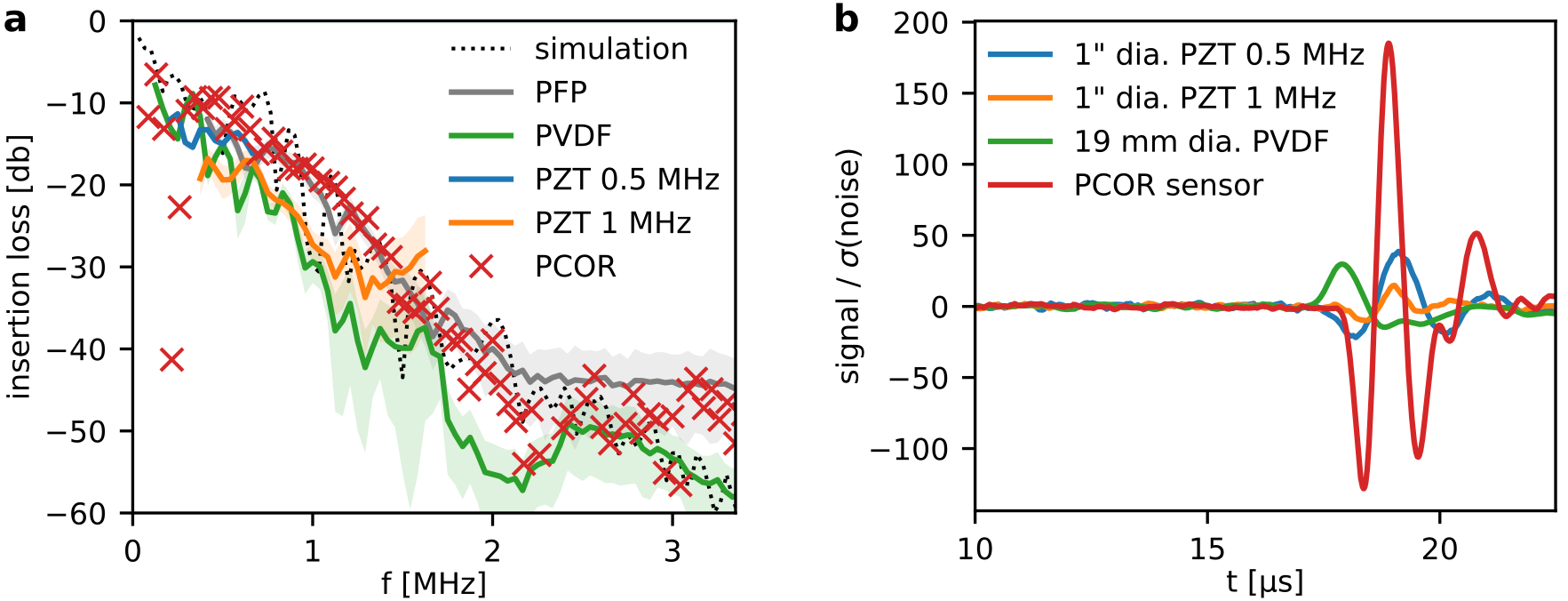}
\caption[insertion losses]{\label{fig:insertionlosses} Frequency spectra and amplitude of PA signals measured through frontal cranial bone using piezoelectric and optical sensors: two PZT transducers with \SI{25.4}{\milli\meter} active element diameter and center frequencies of \SI{0.5}{MHz} and \SI{1.0}{MHz}, one broadband PVDF transducer with \SI{19.2}{\milli\meter} active element diameter, a planar Fabry-Perot (PFP) sensor, and PCOR sensor with a thickness of $L$\,=\,\SI{493}{\micro\meter}. (\textbf{a}) Mean (solid line) and standard deviation (shaded) of the insertion loss of the sensors. The results of the numerical simulation are shown for comparison (dotted line). (\textbf{b}) Representative PA waveforms measured through the frontal cranial bone using the sensors. All waveforms are bandpass filtered and normalized with respect to the detection noise.}
\end{figure}

The frequency dependent insertion loss of \emph{ex vivo} frontal human cranial bone was measured using three piezoelectric transducers with large active element size and two optical sensors, i.e., the planar Fabry-Perot sensor and a plano-concave optical resonator (PCOR) sensor with comparatively small active element sizes (hundreds of microns). Figure \ref{fig:insertionlosses}a shows the insertion loss of the human frontal cranial bone sample measured using the different US sensors. For each sensor, the insertion loss is calculated by dividing the Fourier transform of each waveform transmitted through the frontal bone by the transform of the reference measurement. The measured insertion losses measured by the different US sensors are comparable and consistent, and are also in good agreement with the \emph{in silico} results. The only noticeable difference arises due to the limited frequency responses of the piezoelectric transducers, which do not cover the full acoustic spectrum transmitted through frontal bone.

Figure \ref{fig:insertionlosses}b shows representative PA waveforms measured using the large element size piezoelectric sensors and a PCOR sensor. All waveforms were band-pass filtered (second-order Butterworth filter, cut-on at 10\,kHz, cut-off at 3\,MHz) and normalized with respect to the standard deviation of the noise. Interestingly, the PCOR sensor clearly outperforms the three piezoelectric sensors despite orders of magnitudes difference in their active element size. We believe this is explained by a combination of three factors. First, the large active element size of the PZT and PVDF detectors may result in cancellation effects if heterogeneous acoustic fields, such as those transmitted through bone, are measured. Second, the resonant frequency response of PZT transducers prevents the detection of some of the strongest frequency components of the transcranial PA field, which are found below 0.5\,MHz, thus reducing the signal amplitude. Third, the inherent thermal noise of PZT materials is dependent on the element size, leading to increased noise in large detectors. 

\subsection{Acoustic sensitivity and frequency response of PCOR sensors}
We evaluated a total of ten PCOR sensors, measuring metrics derived form their optical transfer functions, such as full-width half-maximum, fringe visibility, free spectral range, and Q-factor, as well as acoustic parameters, such as the cutoff frequency (or first minimum of the frequency response), the -3\,dB bandwidth, and the minimum and median NEP values. These measured parameters are listed in supplementary table 1. The physical thickness $L$ was calculated from the free spectral range of the sensors. Bandwidth and band-pass cut-off were estimated based on the material parameters and $L$ using a numerical model.\cite{beardTransductionMechanismsFabryPerot1999} The estimated values obtained using the model were in good agreement with measured data as evidenced by supplementary figure 1. The NEP was measured at different output power settings of the interrogation laser (\SI{4}{mW} and \SI{8}{mW}). The NEP did not improve linearly with interrogation laser power as initially expected. Given the high Q-factors of the PCOR sensors, this suggests that at high laser powers phase noise is playing a more dominant role compared with laser intensity noise, thermal noise, and noise from the detection electronics, such as the amplifier.

\begin{figure}[hbt]
\centering
\includegraphics{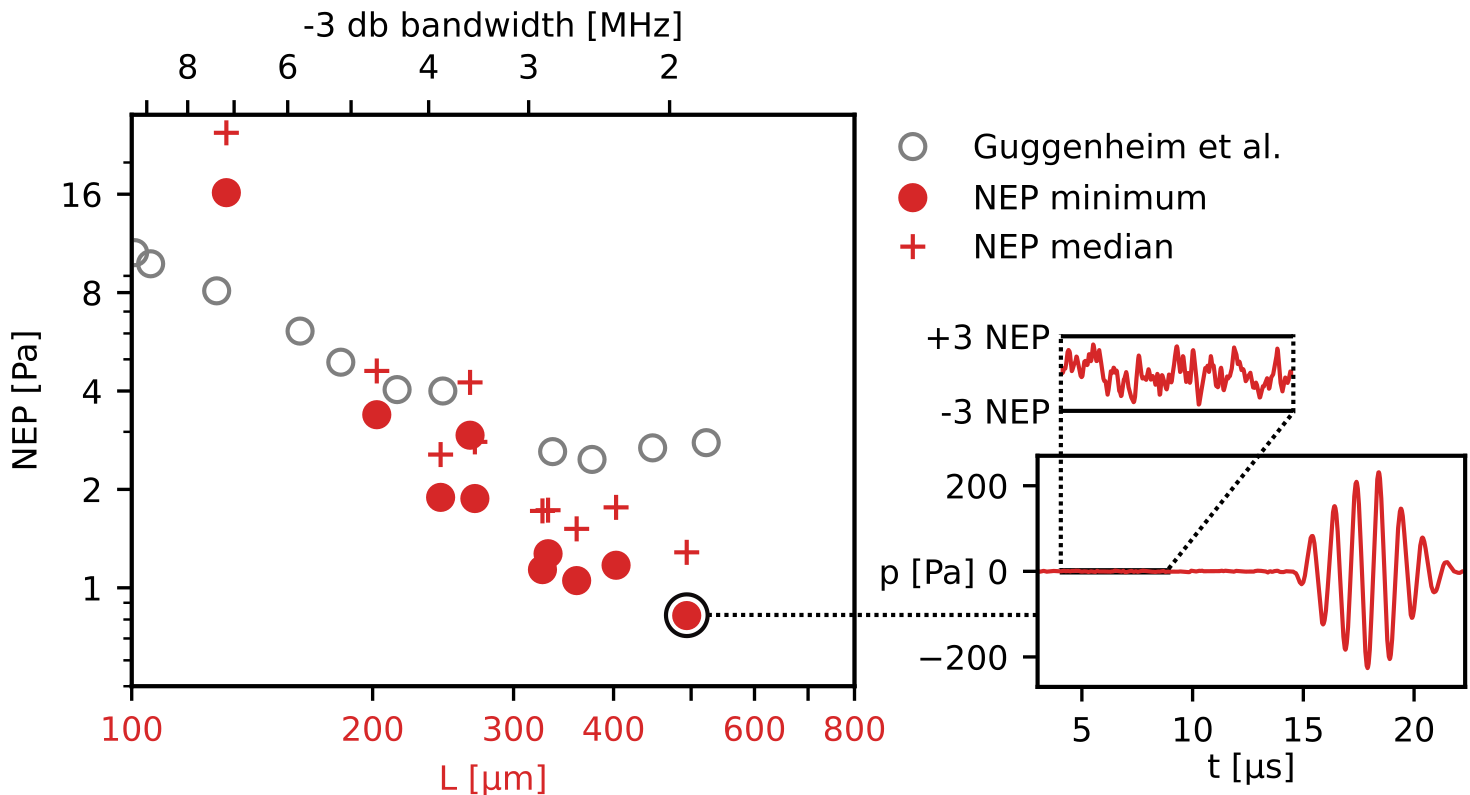}
\caption[NEP]{\label{fig:NEP} Noise equivalent pressure (NEP) of plano-concave optical resonator (PCOR) sensors as a function of sensor bandwidth. The sensitivity of PCOR sensors first reported by Guggenheim et al.~\cite{guggenheimUltrasensitivePlanoconcaveOptical2017} are shown for comparison.  Note that the physical thickness $L$ of the sensors refers to the PCOR sensors developed in this study. Minimum and median NEP were calculated from 100 sequential measurements. The measurement which yielded the lowest NEP of 0.8 Pa is highlighted and the corresponding time-resolved signal is shown.}
\end{figure}

Figure \ref{fig:NEP} shows the NEP of the PCOR sensors as a function of thickness and -3 \,dB bandwidth. Most sensors exhibited NEPs below that reported by Guggenheim et al\cite{guggenheimUltrasensitivePlanoconcaveOptical2017} of \SI{2.6}{Pa}. The largest PCORs produced minimal NEP below \SI{1}{\pascal} (or $\SI{0.6}{\milli\pascal\per\sqrt{\hertz}}$). 
The median NEP was typically above \SI{1}{Pa}. The improvements in sensitivity may be explained by the optical properties of the spacer material (OrmoClad) and the mechanical properties of the substrate material (COP), which has been shown to result in larger acoustic bandwidth\cite{buchmannCharacterizationModelingFabryPerot2017}.
The low optical attenuation of the OrmoClad polymer spacer is perhaps the main reason for the improvement in acoustic sensitivity as it reduces optical losses, leads to sharper resonances in the optical transfer function, and hence an increase in the optical phase sensitivity. Low absorption by the spacer material is particularly beneficial in high-Q, large-L resonators where strong optical confinement results in large path lengths.

The measurements of the optical power reflected by the PCOR sensors as a function of interrogation wavelength, i.e., the interferometer transfer function, were made using a beam waist radius of \SI{30}{\micro\meter}. However, this fixed beam waist and related divergence is not optimal for all PCOR sensors evaluated \cite{martin-sanchezABCDTransferMatrix2023a}. We conducted initial experiments in which the beam waist and divergence was matched to a smaller PCOR and found that fringe visibility was improved (results not shown). This suggests that further improvements in acoustic sensitivity are possible.

We have shown in section \ref{Ex vivo transcranial PA measurements with low-frequency ultrasound sensors} that transcranial PA sensing through thick cranial bone requires US sensors that offer high sensitivity at frequencies below 1\,MHz. The frequency response of the PCOR sensors developed as part of this work is well matched to the transmission spectrum of cranial bone. It is broadband and near-uniform from dc to 1\,MHz. This required bandwidth could even be achieved with PCORs twice the physical thickness of the largest sensors characterised here, leaving room for further improvements in acoustic sensitivity.

The comparison of transcranial PA waveforms shown in figure \ref{fig:insertionlosses}b illustrates that PCOR sensors perform better than much larger conventional piezoceramic sensors, which are generally seen as the most sensitive detectors for PA imaging. When comparing the NEP of PCOR sensors with reported optimized piezocomposite transducers,\cite{xiaOptimizedUltrasoundDetector2013} an at least three orders of magnitude larger active area of piezocomposite is required to match the NEP of the evaluated PCOR sensors. The small element size of PCORs will also lead to less spatial averaging compared to piezoelectric sensors.

\section{Conclusions}
We have shown both \emph{in silico} and in PA transmission measurements through \emph{ex vivo} human skull that ultrasound sensors which have high acoustic sensitivity to low frequency ultrasound are needed to perform transcranial PA measurements in humans. We have illustrated the effects of skull bone tissue on acoustic propagation in PA images acquired using a tomograph based on a broadband planar Fabry-Perot sensor. While the exact frequency dependent losses are dependent upon the type of skull bone, our evaluation of insertion losses have shown that frequencies below 1\,MHz are least attenuated in occipital and frontal cranial bone while temporal bone exhibits a broader acoustic transmission spectrum. We designed, fabricated and characterized an ultrasound sensor based on a plano-concave optical resonator, which combines advantageous attributes for transcranial PA measurements, such as high acoustic sensitivity with NEPs down to \SI{0.8}{Pa} (or $\SI{0.6}{mPa\per\sqrt{Hz}}$), a broadband and near-uniform frequency response, and a small active element size radius of \SI{30}{\micro\meter}.

%%%%%%%%%%%%%%%%%%%%%%%%%%%%%%%%%%%%%%
\subsection*{Disclosures}
The authors have no relevant financial interests in this article and no potential conflicts of interest to disclose.

\subsection*{Acknowledgments}
The authors would like to thank Heike Kielstein, Institute of Anatomy and Cell Biology, MLU Halle-Wittenberg for the provision of the human skull sample; John Maximilian Köhne and Hans-Jörg Vogel, Department of Soil Physics, UFZ for assistance with and access to the X-ray micro-CT machinery;  Werner Lebek of the Medical Physics Department, MLU Halle-Wittenberg for the application of the parylene C coating.

This work was funded by the German Research Foundation (DFG, Deutsche Forschungsgemeinschaft) under grant number 471755457.

\subsection*{Author Contributions}
TK: Conceptualization, Data curation, Formal analysis, Investigation, Funding acquisition, Implementation, Validation, Visualization, Software, Writing -- original draft. 
CV: Methodology and Implementation (for PCOR sensor fabrication). TK and JL:  Methodology, Supervision, Project administration and Resources (also see non-author contributions in acknowledgements). TK, CV and JL: Writing, review \& editing.

\subsection*{Code, Data, and Materials Availability} 
The X-ray Micro-CT data set used in this work is available at \href{https://doi.org/10.5281/zenodo.6108435}{doi:10.5281/zenodo.6108435}. The raw PA and US data is available at \href{https://doi.org/10.5281/zenodo.7998753}{doi:10.5281/zenodo.7998753}.
Data processing and analysis was primarily performed using the open-source k-Wave toolkit as described.

%%%%% References %%%%%

\bibliography{report}   % bibliography data in report.bib
\bibliographystyle{spiejour}   % makes bibtex use spiejour.bst

\renewcommand{\figurename}{Supplementary Figure}
\setcounter{figure}{0}
\renewcommand{\tablename}{Supplementary Table}
\setcounter{figure}{0}
{\footnotesize
\begin{table}[tbh]
\caption{Parameters derived from the interferometer transfer function of ten PCOR sensors and the noise equivalent pressure for an interrogation laser power of 4\,mW and 8\,mW .}
\centering
\begin{tabular}{rrrrrrrrrrr}
\toprule
\multicolumn{1}{r}{\begin{tabular}[r]{@{}c@{}}$L$\\ {[}$\unit{\micro\meter}${]}\end{tabular}} & \multicolumn{1}{r}{\begin{tabular}[r]{@{}c@{}}FWHM\\ {[}$\unit{\pico\meter}${]}\end{tabular}} &\multicolumn{1}{r}{\begin{tabular}[r]{@{}c@{}}visi-\\bility\end{tabular}} & \multicolumn{1}{c}{$Q$} & \multicolumn{1}{r}{\begin{tabular}[r]{@{}c@{}}BP\\ cutoff\\ {[}$\unit{\mega\hertz}${]}\end{tabular}} & \multicolumn{1}{r}{\begin{tabular}[r]{@{}c@{}}-3 db\\ BW\\ {[}$\unit{\mega\hertz}${]}\end{tabular}} & \multicolumn{1}{r}{\begin{tabular}[r]{@{}c@{}}med\\ NEP\\ 4\,mW\\ {[}$\unit{\pascal}${]}\end{tabular}} & \multicolumn{1}{r}{\begin{tabular}[r]{@{}c@{}}med\\ NEP\\ 8\,mW\\ {[}$\unit{\pascal}${]}\end{tabular}} & \multicolumn{1}{r}{\begin{tabular}[r]{@{}c@{}}min\\ NEP\\ 4\,mW\\ {[}$\unit{\pascal}${]}\end{tabular}} & \multicolumn{1}{r}{\begin{tabular}[r]{@{}c@{}}min\\ NEP\\ 8\,mW\\ {[}$\unit{\pascal}${]}\end{tabular}} & \multicolumn{1}{r}{\begin{tabular}[r]{@{}c@{}}min\\ NEP(f)\\ {[}$\frac{\unit{\milli\pascal}}{\sqrt{\unit{Hz}}}${]}\end{tabular}} \\
493 & 27 & 90\% & 60000 & 4.0 & 1.9 & 1.38 & 1.28 & 0.85 & 0.82 & 0.60 \\
403 & 30 & 53\% & 53000 & 4.9 & 2.3 & 2.59 & 1.76 & 1.80 & 1.17 & 0.77 \\
360 & 21 & 60\% & 74000 & 5.5 & 2.6 & 2.15 & 1.51 & 1.45 & 1.05 & 0.65 \\
331 & 25 & 61\% & 63000 & 6.0 & 2.8 & 2.10 & 1.73 & 1.41 & 1.27 & 0.75 \\
326 & 25 & 54\% & 64000 & 6.1 & 2.9 & 2.45 & 1.72 & 1.59 & 1.14 & 0.67 \\
268 & 34 & 46\% & 46000 & 7.4 & 3.5 & 3.22 & 2.80 & 2.35 & 1.88 & 1.00 \\
265 & 42 & 40\% & 38000 & 7.5 & 3.6 & 4.40 & 4.25 & 3.07 & 2.93 & 1.55 \\
243 & 33 & 56\% & 48000 & 8.1 & 3.9 & 3.09 & 2.56 & 2.14 & 1.89 & 0.96 \\
202 & 44 & 50\% & 36000 & 9.8 & 4.6 & 6.06 & 4.61 & 4.57 & 3.38 & 1.57 \\
131 & 51 & 47\% & 31000 & 15.1 & 7.2 & 37.86 & 24.65 & 18.91 & 16.17 & 6.04 \\
\bottomrule
\end{tabular}
\end{table}
}

\begin{figure}[hbt]
\centering
\includegraphics{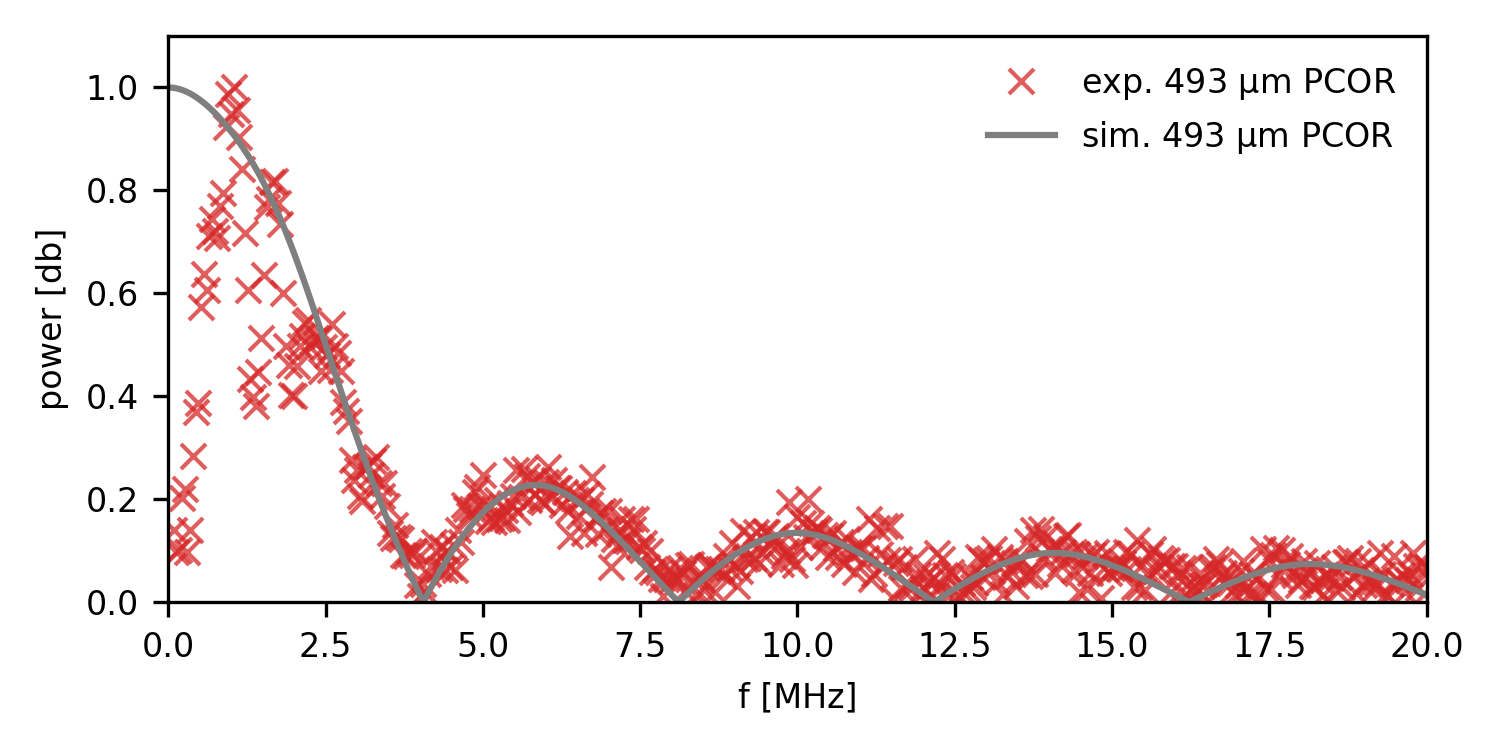}
\caption[PCOR frequency response]{\label{fig:PCORfreq} Comparison between the simulated and experimental frequency response of a plano-concave optical resonator (PCOR) sensor with a physical thickness of \SI{493}{\micro\meter}.}
\end{figure}
{\footnotesize
\noindent\textbf{Supplementary Video 1} -- \url{https://bit.ly/transcranialPA} -- Acoustic propagation through frontal cranial bone \emph{in silico}. A 3D k-Wave forward simulation of a flat acoustic source with 5\,mm diameter. Time series of the pressure distribution in the center slice. The light gray area shows the skull segmentation used for the simulation.
}

\end{spacing}
\end{document}